\documentstyle[prd,aps,psfig]{revtex} 

\newcommand\be{\begin{equation}}  \newcommand\ee{\end{equation}}
\newcommand\bea{\begin{eqnarray}}  \newcommand\eea{\end{eqnarray}}
\newcommand\ba{\begin{array}}  \newcommand\ea{\end{array}}
\begin{document}
\draft \input epsf  \def\la{\mathrel{\mathpalette\fun <}}
\def\ga{\mathrel{\mathpalette\fun >}}
\def\fun#1#2{\lower3.6st\vbox{\baselineskip0st\lineskip.9st
\ialign{$\mathsurround=0pt#1\hfill##\hfil$\crcr#2\crcr\sim\crcr}}}

\twocolumn[\hsize\textwidth\columnwidth\hsize\csname
@twocolumnfalse\endcsname

\title{Anarchy in the neutrino sector?}
\author{J.R. Espinosa} 
\address{\phantom{ll}}
\address{{\it IMAFF, CSIC, Serrano 113 bis, 28006 Madrid, Spain}} 
\address{{\it IFT, C-XVI, Univ. Aut\'onoma de Madrid, 28049 Madrid, 
Spain}}
\date{\today}  \maketitle
\begin{abstract}
If the connection between the fundamental theory and neutrino mass 
matrices $M_\nu$ is sufficiently complicated all traces of underlying 
symmetries might be erased and an 'anarchical'  $M_\nu$, with random 
entries, can arise. It has been claimed that such random matrices 
generically 
prefer large mixing making unnecessary to invoke flavour 
symmetries to explain the observed pattern of neutrino mixing angles. A 
critical analysis 
of this idea (paying particular attention to the r\^{o}le of unphysical 
phases and coordinate dependence) shows that basis-independent, random  
$M_\nu$'s are not biased towards large values of the 
neutrino mixing angles, contrary to previous claims. As a result, the case 
for anarchy in the neutrino sector is considerably weakened.
 \end{abstract}
\pacs{PACS: 14.60.Pq, 12.60.Jv~ 
IFT-UAM/CSIC-03-18.
\hspace*{4cm}  [hep-ph/0306019]} ]


{\bf 1. Introduction.} In a series of three interesting papers
\cite{I,II,III} the idea that the measured neutrino parameters
\cite{concha} could arise from a random mass matrix has been proposed and
studied. A neutrino mass matrix with random entries could be the result of
a fundamental theory that is itself complicated or that requires a
convoluted mechanism ({\it e.g.} with several sources) for generating 
neutrino masses.

In ref.~\cite{I}, a random scan of $10^{6}$ neutrino mass matrices (of Dirac, 
Majorana or see-saw type) is performed and the resulting distributions 
for mixing angles and for the ratio $R\equiv\Delta m^2_{sol}/\Delta 
m^2_{atm}$ are studied. The distributions for the atmospheric and solar 
angles peak at maximal mixing (which is welcome) while $U_{e3}$ is in 
general not particularly small (which can be a problem, or a promise for 
its measurement in the future). Finally, it is easy to get a small $R$ 
($\sim 0.1$), especially in the see-saw case.

Ref.~\cite{II} sheds lihgt on the previous results by realizing that the
distributions of mixing angles are in fact dictated by the requirement of
basis independence of the random scan. 
If $M_\nu$ is
diagonalized by a mixing matrix $U$ that belongs to a certain Lie group
[{\it e.g.} $U(3)$ for a complex $3\times 3$ Majorana $M_\nu$], then the
probability distribution for the mixing angles that parametrize $U$ follow
the Haar measure of that Lie group. This nice property allows an analytic
understanding of the predictions of anarchy.

Finally, ref.~\cite{III} tests the hypothesis of anarchy in the neutrino sector 
by performing a Kolmogorov-Smirnov (KS) test taking as data the observed 
neutrino mixing angles (choosing the best fit points for the solar and 
atmospheric mixing angles and leaving the CHOOZ angle free, but below its 
bound). The test should measure how well the anarchy hypothesis 
describes the data or, alternatively, see if the data are able to rule 
out this hypothesis (with some degree of C.L.). It is found that the KS 
probability in favor of the anarchy hypothesis is $\sim 12\%$.

In this letter, the predictions of the anarchy hypothesis for different
types of neutrino mass matrices are carefully re-examined in section~2.
This analysis shows that anarchy has no preference for certain ranges of
mixing angles. In particular, there is no preference for maximal mixing. A
critical analysis of statistical tests of the anarchy hypothesis,
performed in section~3, confirms this result.

{\bf 2. Mixing angles predicted by Anarchy.}
One of the main ingredients of the hypothesis of anarchy in the neutrino 
sector is the requirement of basis-independence in flavour space: {\it 
i.e.} the entries of $M_\nu$ are random numbers 
distributed in some specified range according to a flat probability 
density, that remains flat in any basis. As an example,
take a $2\times 2$ real and symmetric $M_\nu$. Factoring out the 
absolute neutrino mass scale $m$, take the entries of ${\hat M}_\nu=
M_\nu/m$ as random numbers in the interval $[-1,1]$ with uniform 
probability density. The differential probability density for each 
entry is simply $d{\hat M}_{ij}$ and in the 3-D space $\{{\hat 
M}_{11}, {\hat M}_{12},{\hat M}_{22}\}$ that parametrizes ${\hat 
M}_\nu$, the differential probability density is given by the 3-form
$dP=d{\hat M}_{11}\wedge d{\hat M}_{12}\wedge d{\hat M}_{22}$.
It is straightforward to prove the invariance of $dP$ under rotations in 
flavour space. (This result can be extended to more general matrices: for 
complex entries one simply uses $d {\hat M}_{ij}\equiv d \Re e~{\hat 
M}_{ij}\wedge d \Im m~{\hat M}_{ij}$.) A uniform probability density for 
the entries of ${\hat M}_\nu$ is not enough to guarantee 
basis-independence. The range in which these entries can vary must also be 
a basis-independent domain. In the previous example, this domain was a 
cube, which is not rotation-invariant, but one can simply
take  ${\hat M}_{ij}$ inside a sphere of some specified radius:
${\mathrm Tr}~{\hat M}_\nu^\dagger {\hat M}_\nu<r^2$.

Once the basis-independence of the random matrices is enforced, the mixing 
matrices $U$ that diagonalize them are distributed in a well determined 
way: if the $U$'s are matrices of a given Lie group, the differential 
probability density for $U$, $d\Omega$, is simply the invariant Haar 
measure of that Lie group (see \cite{II}). In terms of the 
invariant Maurer-Cartan 
1-forms $\omega^a\equiv-i{\mathrm Tr}~T^aU^{-1}dU$, where $T^a$ are the 
generators of the Lie group, then 
$d\Omega=\epsilon_{a_1,a_2,...a_N}\omega^{a_1}\wedge ...\wedge 
\omega^{a_N}$, with $N$ the dimension of the group.

Let us now examine in turn several examples of Majorana mass matrices 
(the results can be extended to Dirac or see-saw matrices).

{\bf 2.1 Real ${\mathbf 2}{\mathbf\times} {\mathbf2}$ 
${\mathbf M}_{\mathbf \nu}$}

These matrices are diagonalized by rotations, $U\equiv R(\theta)$, so that
the corresponding Lie group is $U(1)$ and $d\Omega=d\theta$, which is
explicitly rotation invariant. The distribution of the mixing angle
$\theta$ is therefore flat and no mixing angle is preferred. The
conclusions of ref.~\cite{II} are different: by writing
$d\theta=d\sin^22\theta/(2\sin4\theta)$ it is shown that the distribution
of $\sin^22\theta$ peaks at zero and maximal angles, which are claimed to
be preferred. Such peaks, however, {\it do not} imply that angles near
zero or maximal mixing are favored: these peaks are simply the result of
using a particular coordinate in the circle, $\sin^22\theta$, which
distorts the uniform distribution in angles. To see this even more
clearly, notice that we have not specified in what basis we are measuring
$\theta$. We would find peaks for $\sin^22\theta$ in any basis, but if
they were to imply a preferrence for maximal and zero mixing, one arrives
at the paradoxical conclusion that different physical angles are preferred
depending on the basis used.

Given the flat distribution in $\theta$, all one can say is that large 
mixing angles are not unlikely. The relative likelihood of finding  
$\theta$ in a region near zero mixing or in a region near maximal mixing 
is simply given by the relative size of the two regions. In section 3 we 
discuss the statistical tests that might be applied to 
data in order to see how well they support the anarchy hypothesis.

{\bf 2.2 Complex ${\mathbf 2}{\mathbf\times} {\mathbf2}$ 
${\mathbf M}_{\mathbf\nu}$}

These Majorana matrices are diagonalized by $U(2)$ matrices that can be 
written as $U=\exp [i(\eta I_2+\omega \sigma_3)]R(\theta)
\exp [i\phi\sigma_3]$, where $\eta$ and $\omega$ are unphysical phases,
$\theta$ is the rotation angle and $\phi$ a $CP$ violating phase. The 
distributions of these quantities in the anarchy scenario follow from the 
$U(2)$ Haar measure
\be
\label{omegau2}
d\Omega=ds_\theta^2\wedge d\eta\wedge d\omega \wedge d\phi\ ,
\ee
(with $s_x\equiv \sin x$).
Rewriting $ds_\theta^2=s_{2\theta} d\theta$, ref.~\cite{II} concludes that, 
in this case, maximal mixing is preferred. This conclusion, which seems 
now inescapable, leaves still the paradox that the preferred angle seems 
to depend on the basis chosen to measure $\theta$ (in spite of the fact 
that $d\Omega$ is indeed basis independent). One might argue that the 
measured $\theta$ is defined in the flavour eigenstate basis and therefore 
such basis should be used. However, one should be allowed to 
determine the probability distributions and the most probable 
angle in a rotated basis, provided one rotates back later to the flavour 
eigenstate basis. The paradox is that, doing so, the final result for the 
expected $\theta$ depends on the intermediate rotated basis used.

In order to understand why, according to (\ref{omegau2}),  $\theta=\pi/4$
seems to be preferred while $\theta=0$ is disfavoured, it is useful to
inspect the explicit expressions for the invariant Maurer-Cartan forms:
\be
\begin{array}{ll}
\omega^1= d\eta\ , & \omega^2= s_{2\theta}c_{2\phi}d\omega
-s_{2\phi}d\theta\ ,\nonumber\\
\omega^3=s_{2\theta}s_{2\phi}d\omega
+c_{2\phi}d\theta\ , & \omega^4= c_{2\theta}d\omega+d\phi\ ,
\end{array}
\label{MComplex}
\ee
with $d\Omega=\omega^1\wedge\omega^2\wedge\omega^3
\wedge\omega^4$. We see that, for $\theta\rightarrow 0$ one has
$\omega^2\wedge\omega^3\rightarrow 0$, while this is not the case for
finite values of $\theta$ due to the presence of $d\omega$ terms. In other
words, the subspace of parameter space corresponding to $\theta=0$ has one
dimension less than that for $\theta=\theta_0\neq 0$ and the extra volume
associated with the latter case opens up in the direction of the
unphysical parameter $\omega$. However, all matrices related to each other
by changes in an unphysical parameter should be considered as physically
indistinguishable and therefore the preference for maximal mixings does
not stand on firm ground.

Keeping the unphysical phases $\eta, \omega$ helps dealing with basis
independence but introduces a bias in the distributions of physical
parameters. How do we get rid of this problem? 
We simply fix $\eta=\omega=0$, which we are free to do for any $M_\nu$. 
The crucial point is that this condition is rotation invariant, as can be
readily checked.  After fixing this `gauge', $\theta$ transforms under a
rotation by angle $\alpha$ simply by $\theta\rightarrow \theta+\alpha$
while the phase $\phi$ does not change (notice that in the presence of
non-zero unphysical phases the transformation properties under rotations
are more complicated). Therefore, after 'gauge-fixing', the only volume
element in the space $\{\theta,\phi\}$ that is rotation invariant must be
of the form\cite{footnote1}
\be
d\Omega_{gf}={\cal F}(\phi)d\theta\wedge d\phi\ .
\label{dOmegac}
\ee 

Because invariance under rotations is not enough to fix ${\cal F}(\phi)$ 
in (\ref{dOmegac}), further assumptions besides anarchy are necessary to 
make
any prediction about the distribution of the phase $\phi$. This situation
is similar to that of the distribution of mass eigenvalues \cite{II}
which are unknown functions of mass differences. In analogy with that
case, one might choose these functions in such a way as to get a volume
element in parameter space that is simple when expressed in terms of mass
matrix elements. Such choice is useful to generate samples of random
matrices like in \cite{I} but now in a 'fixed gauge'; however, its
theoretical motivation is not clear. A simple choice for a rotation
invariant volume element (in terms of matrix entries) is
$d V_{gf}=d\Re e M_{11}\wedge d\Re e M_{22}\wedge d\Re e M_{12}\wedge
d\log\Im m M_{12}$, which leads to
\be
d V_{gf}=-2(m_1-m_2)dm_1\wedge dm_2\wedge c_\phi^2d\phi\wedge d\theta
\ .
\ee
This privileges small values of $\phi$ but, as explained, other choices of
invariant volume are possible.  The uncertainty in the distribution of
$\phi$ does not affect the distribution of $\theta$, which is again flat:
all mixing angles are equally probable and the comments at the end of
section~2.1 apply also to this case.

{\bf 2.3 Real ${\mathbf 3}{\mathbf\times} {\mathbf3}$ 
${\mathbf M}_{\mathbf \nu}$}

A symmetric real $M_\nu$ can be diagonalized by an orthogonal matrix $U$
belonging to $SO(3)$. The parametrization
$U=R_{23}(\theta_1)R_{31}(\theta_2)R_{12}(\theta_3)$, with
$R_{ij}(\theta)$ a rotation in the plane $ij$ by an angle $\theta$, is
quite convenient to discuss neutrino oscillations. The angle $\theta_3$ is
the relevant angle for the oscillations of solar neutrinos, $\theta_2$ is
the small mixing angle bounded by CHOOZ experiment and $\theta_1$ is the
relevant angle for the oscillations of atmospheric neutrinos 
\cite{concha}. The
differential probability distributions for the mixing angles $\theta_i$
derived from anarchy is then dictated by the $SO(3)$ invariant Haar
measure
\be
d\Omega=d\theta_1\wedge \cos\theta_2 d\theta_2\wedge d\theta_3\ .
\label{SO3}
\ee
In order to understand
better what are the implications of (\ref{SO3}) concerning the likelihood
of different values for the mixing angles, it is convenient to switch to a
different parametrization of $U$. Let $\vec{n}$ be the axis of rotation in
flavour space [parametrized in terms of polar angles ${\alpha,\beta}$ as
$\vec{n}=(s_\beta c_\alpha,s_\beta s_\alpha,c_\beta)$], and $\theta$ the
rotation angle. We can then write
$U=R_{12}(\alpha)R_{31}(\beta)R_{12}(\theta)R_{31}(-\beta)R_{12}(-\alpha)$.
In terms of these parameters one obtains
\be
d\Omega=d\alpha\wedge \sin\beta d\beta \wedge \sin^2(\theta/2)d\theta\ .
\label{SO3p}   
\ee

For a fixed value of $\theta$ this measure is just that on the sphere
($d\alpha\wedge \sin\beta d\beta$) with polar coordinates $\alpha,\beta$.
Although this measure is zero for $\sin\beta=0$, this does not mean that
rotations along the $z$ axis are disfavoured but rather that the
distribution of the rotation axis $\vec{n}$ is uniform in flavour space,
as it should be the case if one starts with a rotation-invariant mass
matrix. A useful analogy is to consider a uniform distribution of points
over the surface of the Earth. In terms of longitude ($\alpha$) and
latitude ($\beta$) the probability density has the same form as above and
the expected number of points inside a certain region is simply
proportional to its area no matter where this region is on the surface of
the Earth.

The previous discussion shows that in order to make meaningful statements
about what angles are preferred, one should take into account the area (or
volume) of the regions of parameter space that are compared. What anarchy
implies, as encoded in (\ref{SO3}), is that all regions of parameter space are
equally likely. The 'preference' for small $\theta_2$ in (\ref{SO3})  is due
to the fact that the volume of parameter space with small $\theta_2$ is larger
[in the same sense that $\beta\sim 0,\pi/2$ is disfavored in (\ref{SO3p})
because the area of the poles is smaller than the rest of the Earth's area].

{\bf 2.4 Complex ${\mathbf 3}{\mathbf\times} {\mathbf3}$ 
${\mathbf M}_{\mathbf \nu}$}

A complex symmetric $M_\nu$ is diagonalized by a $U(3)$ 
matrix that may be written in this way:
$U={\mathrm diag}(e^{i\alpha_1},e^{i\alpha_2},e^{i\alpha_3})$ 
$R_{23}(\theta_1)$$e^{-i\delta \lambda'}$$R_{13}(\theta_2)$
$e^{i\delta\lambda'}$$R_{12}(\theta_3)$
${\mathrm diag}(e^{-i(\xi_1+\xi_2)},e^{i\xi_1},e^{i\xi_2})$
with $\lambda'={\mathrm diag}(1,0,-1)/2$. In this notation $\alpha_{1,2,3}$ 
are  
unphysical phases. The distribution of angles and phases according 
to the original idea of anarchy would then follow the distribution
\[
d\Omega=ds_3^2\wedge dc_2^4\wedge
ds_1^2\wedge d\delta\wedge d\xi_1\wedge d\xi_2\wedge d\alpha_1 
\wedge d\alpha_2\wedge d\alpha_3
\]
which corresponds to the $U(3)$ Haar measure. This results from combining
9 invariant Maurer-Cartan forms (straightforward to compute but
lengthy).

As was discussed for the complex $2\times 2$ case, one cannot use
$d\Omega$ above to extract physical implications without getting rid of
the unphysical degrees of freedom (which otherwise affect the distribution
of physical parameters). This problem is solved by 'choosing a gauge' like
$\alpha_1=\alpha_2+\delta$ and $\alpha_2=\alpha_3=-\xi_2$, which is
invariant under rotations of basis. 
The invariant volume element after this
gauge-fixing is of the form \cite{footnote3}
\bea
d\Omega_{gf}&=&{\cal F}(\delta-3\xi_2){\cal G}(s_\delta 
s_3c_3)(c_2d\theta_1\wedge
d\theta_2\wedge d\theta_3)\nonumber\\
&&\wedge (d\xi_1\wedge d\xi_2\wedge d\delta/s_\delta)\ .
\label{omegagf}
\eea
This volume element is composed of three parts: ${\cal F}$ and ${\cal G}$ 
are unknown 
functions of their invariant arguments; the angular element is the volume 
element for real $3\times 3$ matrices (\ref{SO3}), which corresponds to a 
homogeneous distribution of mixing angles; and finally 
there is a phase volume element. The predictions of anarchy for the angles 
$\theta_1$ and $\theta_2$ are similar to those of the real $3\times 3$ 
case: all values of $\theta_1$ are equally likely; small $\theta_2$ is 
preferred due to volume effects. Without further assumptions on the form 
of ${\cal F}$ and ${\cal G}$, anarchy does not make more detailed 
predictions for 
$\theta_3$ and the physical phases (except for  the prediction of 
a flat distribution for $\xi_1$). If ${\cal G}(x)=x$, $d\Omega_{gf}$ gives 
a 
preference for large solar mixing.

{\bf 3. Statistical test of the hypothesis of anarchy}

In order to assess how likely it is that the observed neutrino angles 
arise from a distribution dictated by the hypothesis of anarchy, 
ref.~\cite{III} performs an statistical test as follows. Consider for 
simplicity the solar angle $\theta_3$ and assume that
 the probability density is flat in $x=\sin^2\theta_3$ (as
was the prediction for $3\times 3$ complex Majorana matrices in
ref.~\cite{III}, revised in this paper) and take $x$ as a random variable
in the interval $[0,1]$ (the same analysis applies to a flat distribution
in $d\theta$, with $x=2\theta/\pi$). Experimentally, the best fit point 
for solar
oscillations corresponds to $x_e=0.3$. Based on this, ref.~\cite{III}
tests the anarchy hypothesis using standard methods, by comparing the
distribution function predicted by anarchy, $F(x)=\int_0^x f(x')dx'=x$,
with the 'empirical' distribution function, $F_e(x)$, obtained from the
best guess probability density, $f_e(x')=\delta(x'-x_e)$, that leads to
$F_e(x)=\int_0^x f_e(x')dx'=\Theta(x-x_e)$. The mismatch between the two
distribution functions is measured by the (two-sided) Kolmogorov-Smirnov
statistic
\be
D=max_x[|F_e(x)-F(x)|]\ .
\label{KS}
\ee
The higher $D$ is, the worst $F(x)$ accomodates the observed data. More 
precisely, the probability of obtaining a worse value of $D$ from a 
different observed $x$ is 
\be
P(D\geq D_e)=\left\{\begin{array}{ll}
2x_e\ , & {\mathrm if}\ x_e\leq 1/2 \\
2(1-x_e)\ , & {\mathrm if}\ x_e\geq 1/2 
\end{array}\right.\ 
\label{kolmo}
\ee
where $D_e$ is the KS statistic for $x_e$. It is clear that, according to 
this procedure, the value $x_e=1/2$ ({\it i.e.} maximal mixing) is the 
best possible result in favor of the anarchy hypothesis. 

This is a puzzling result.  How can a single data point be able to provide
evidence in favor of a flat distribution at all? Why should data values
close to $x_e=1/2$ be the best support for a flat probability distribution
for $x$?  While this seems reasonable for $x=\sin^2\theta_3$ it is
certainly not for $x=2\theta/\pi$, and ref.~\cite{III} uses the above
test, without modification, also for distributions flat in
$x=\cos^4\theta$. There are three problems with the previous analysis.

The first is that $x$ is an angular variable ($0\leq\theta<\pi/2$ is mapped 
into the interval $[0,1]$ with $x=0$ and $x=1$ corresponding to the same 
physical point) and the
test used is suited for linear variables. In fact the statistical analysis of
angular or directional data requires some departure from the usual linear
statistical methods and there is a well developed corpus of knowledge on this
topic (for an introduction to this subject see {\it e.g.} \cite{circstat})
which is relevant in many areas of science, from the analysis of directional
patterns in bird migration to magnetic pole wandering, coincidence of
planetary orbital planes or distribution of earthquakes in the globe, among
others.  The version of the Kolmogorov statistic that is best suited to
angular random variables is Kuiper's test \cite{circstat}, defined as
\cite{footnote2}
\be
D=max_\theta[F_e(\theta)-F(\theta)]+max_\theta [F(\theta)-F_e(\theta)]\ .
\label{K}
\ee
This statistic has the desirable property of being independent of the 
choice of origin for $\theta$ (see Appendix for a simple proof), which is 
not the case for the linear Kolmogorov statistic, Eq.~(\ref{KS}). 

However, the previous improvement is not sufficient in view of the second
problem: the smallness of the sample. One can easily see that the Kuiper
statistic for a single random draw, $\theta_e$, is $D_e=1$, which is
independent of $x_e$. Although this seems to be a reasonable result (it
implies that a single data point cannot give evidence in favor or against the
hypothesis of a flat probability distribution) it cannot be taken 
seriously because the Kuiper test is always $D_e=1$ for a single data point
irrespective of the distribution being tested! One possible way out would be
to improve the test by taking into account the difference between the
distribution functions over the whole interval ('integral' variants of the
Kuiper test exist \cite{Watson}) but, as we show below, these improvements 
would also be insufficient. 

The third problem is that the K-S test is invariant under
reparametrizations of $x$, {\it e.g.} it cannot tell apart distributions
that are flat in $x=\sin^2\theta$ from those that are flat in 
$x=2\theta/\pi$, and this is inappropriate.

An alternative approach (for $x=2\theta/\pi$) is to enlarge the data
sample by considering the origin, $\theta_0=0$, defined by the flavour
basis, as another data point. In that case the probability of getting a
fit worse than the observed one, according to the Kuiper test, is given
again by Eq.~(\ref{kolmo}) with $x_e=2(\theta_e-\theta_0)/\pi$, which is
explicitly rotation invariant. From this result one would conclude that
$\theta_e-\theta_0\sim \pi/4$ would give the strongest support to a flat
distribution and $\theta_e-\theta_0\sim 0, \pi/2$ the weakest, which
agrees with the conclusions of Ref.~\cite{III}. However, this result
cannot be trusted. The reason is that even two points are too few to give
evidence in favor of a flat distribution function: any set of two data
points has the same probability of any other (this situation changes only
for three or more data points).

We therefore abandon the attempt at performing a meaningful statistical
test of the anarchy hypothesis, at least for uniformly distributed angles
(in principle a partial test could be devised for angles which are not
uniformly distributed, like $\theta_2$) and prefer to limit ourselves to
the following estimate, similar to the analysis performed in
Ref.~\cite{I}.  Assuming a large number of random neutrino mass matrices,
we can estimate now what fraction of them will have angles in some
neighbourhood of the observed values (say in a subspace $v$ of the
total volume $V$).  For $3\times 3$ real Majorana matrices [or for complex
ones with ${\cal G}(x)=1$ in Eq.~(\ref{omegagf})], that fraction will be 
given
simply by the ratio $\int_v d\Omega/\int_V d\Omega$.
Using ranges similar to those in
Ref.~\cite{I} ($\sin^22\theta_2<0.15$, $\sin^22\theta_{1,3}>0.5$) the
interesting area is about a $4.8\%$ of the total. This number is similar
to the one obtained in Ref.~\cite{I} for the real case (in which case we
both agree on the angle distributions) before applying the cut in mass
splittings.  If ${\cal G}(x)=x$ is chosen in (\ref{omegagf}), then the 
preference
for maximal solar mixing raises the previous percentage to 6.8\%. Still,
we regard this as a very small basis upon which to build the case in
favour of anarchy. In contrast, attempts based on flavour symmetries fare
considerably better \cite{Hirsch,AFM} and are physically much more
appealing: they imply that what we learn at low-energy from measurements
of the parameters in the neutrino sector will be a window (with a view) to
fundamental physics at much higher energies.

{\bf Appendix}
To see that the Kuiper's test is invariant under rotations,
note first that, due to the 'periodicity' property of angular distribution 
functions, $[F(\theta+2\pi)=F(\theta)+1]$, any angular interval 
$[\alpha,\alpha+2\pi]$ is suitable to compute $D$ in Eq.(\ref{K}).
If we now change from the random variable $\theta$ to the rotated
$\theta'=\theta+\alpha$, the new distribution functions are
$F'(\theta')=F(\theta+\alpha)-F(\alpha)$ and
\bea
D'&=& max_{\theta'}[F'_e(\theta')-F'(\theta')]+max_{\theta'} 
[F'(\theta')-F'_e(\theta')]\nonumber\\
&=& 
max_{\theta}[F_e(\theta+\alpha)-F(\theta+\alpha)]-
F_e(\alpha)+F(\alpha)\nonumber\\
&+&max_{\theta}[F(\theta+\alpha)-F_e(\theta+\alpha)]+F_e(\alpha)-F(\alpha)
\nonumber\\
&=&max_{\theta}[F_e(\theta)-F(\theta)]+
max_{\theta}[F(\theta)-F_e(\theta)]\nonumber\ ,
\eea
so that $D'=D$. A less straightforward proof of this property can be 
found in \cite{circstat}.

{\bf Acknowledgments:} Useful discussions with 
Guido Altarelli, Alberto Casas, Concha 
Gonz\'alez-Garc\'{\i}a, Alejandro Ibarra and Ra\'ul Rabad\'an 
are gratefully acknowledged. I also thank CERN, where most of
this work was carried out, for hospitality and partial financial support.


\begin{references}
\bibitem{footnote1}
The same result can be obtained setting $d\eta=d\omega=0$ in the
Maurer-Cartan forms (\ref{MComplex}). One gets the three invariant forms
$s_{2\phi}d\theta$, $c_{2\phi}d\theta$ and $d\phi$. Invariance of the
first two forms implies invariance of $\phi$ and $d\theta$ separately, and
the combination of these forms leads to Eq.~(\ref{dOmegac}).
\bibitem{footnote3}
This can be checked either by direct
calculation or by setting $d\alpha_1=d\alpha_2+d\delta$ and
$d\alpha_2=d\alpha_3=-d\xi_2$ in the $U(3)$ Maurer-Cartan forms to
identify invariant zero- and one-forms out of which $d\Omega_{gf}$
is constructed.
\bibitem{footnote2}
Distribution functions for an angular variable $\theta$ are simply 
defined by \cite{circstat} $F(\theta)=\int_0^\theta f(\theta')d\theta'$
for $0<\theta\leq 2\pi$ and extended to all values of $\theta$ by
$F(\theta+2\pi)=F(\theta)+1$.
\end{references}
\end{document}